# Apport des mesures de champ dans l'étude de composites renforcés en fibres de lin au cours d'essais de traction quasi-statiques

*Contribution of field measurements in the study of composites reinforced with flax fibers during quasi-static tensile tests*

**Amélie Cuynet[1]\*, Franck Toussaint[1], Emile Roux[1], Daniel Scida[2], Rezak Ayad[2]**

[1] *Laboratoire SYstèmes et Matériaux pour la MEcatronique (SYMME)*
*Université Savoie Mont Blanc, France*
\* Auteur correspondant : amelie.cuynet@univ-smb.fr

[2] *Laboratoire d'Ingénierie et Science des Matériaux (LISM)*
*Université de Reims Champagne-Ardenne, France*

### Résumé

Ce travail porte sur l'étude du comportement de matériaux composites constitués d'une résine époxyde et de tissus à fibres de lin à renfort sergé. Le but de cette étude est de déterminer les module d'élasticité, coefficients de Poisson et contrainte à la rupture, de ce type de matériau par corrélation d'images. Les échantillons utilisés dans le cadre de ce travail ont été fabriqués par le procédé de moulage par injection sous vide. Ils ont ensuite été soumis à des essais de traction quasi-statiques. Deux appareils-photos ont été placés perpendiculairement à la face et la tranche de l'échantillon de manière à obtenir une séquence d'images numériques visibles au cours de l'essai. Les images enregistrées ont été analysées avec le logiciel de corrélation d'images *7D* afin de calculer les champs de déplacements et déformations *ad hoc*. Afin de vérifier la validité des résultats de l'analyse, ces mesures sont comparées à celles obtenues par extensométrie classique. On montre que malgré les très faibles niveaux de déformation, les résultats des identifications issus de l'extensométrie optique sont très prometteurs.

### Abstract

This work is based on the study of the behaviour of flax fibres twill reinforced epoxy composites. The aim of the study is to determine the modulus of elasticity, Poisson coefficients and strength of these composites materials with image correlation. The specimens used in this work were manufactured by the infusion vacuum process and after subjected to quasi-static tensile tests. Two numeric cameras were disposed perpendicular to the front and to the side of the specimen so as to achieve a sequence of visible digital pictures during the test. The recorded pictures were analysed with the image correlation software 7D in order to calculate ad hoc measurements of displacements and strain fields. To check the validity of the results, measurements were compared to those obtained by classical extensometry. We found that, despite the low levels of strains, results of identifications from optical extensometry were very promising.

**Mots Clés :** composites à fibres de lin, corrélation d'images, propriétés mécaniques, essais de traction.
**Keywords :** flax-fibre composites, digital image correlation, mechanical properties, tensile test.

## 1. Introduction

Ce travail s'intègre dans le projet de recherche Bioimpact et a pour but d'étudier le comportement mécanique de matériaux composites à fibres de lin. Ces matériaux sont particulièrement intéressants grâce aux spécificités des fibres de lin puisque certaines de ses propriétés se rapprochent de celles de la fibre de verre. De plus, étant beaucoup plus légère que la fibre de verre, la fibre de lin constitue en fait une très bonne alternative pour remplacer les fibres de verre. Plusieurs études ont déjà été réalisées lors d'essais statiques, et en particulier pour des composites à renfort sergé à fibres de lin [1]. Le projet Bioimpact vise entre autres à reprendre ces matériaux pour avancer dans leur connaissance, l'objectif étant d'analyser leurs modes de rupture, notamment par rapport à la vitesse de sollicitation. Il s'agit ici de présenter une étude préliminaire à ce projet sur l'apport des mesures de champs de déformation, mesures réalisées dans le cadre d'une sollicitation simple de traction. Les échantillons sont composés d'une matrice époxyde et de renforts tissés à fibres de lin, dont l'armure est un sergé 2/2. A partir d'essais de traction, les déformations de l'éprouvette sont obtenues à l'aide d'une technique de corrélation d'images numériques visibles utilisant deux appareils-photos Nikon

D500. L'un des appareils prend des images de la tranche de l'éprouvette et l'autre prend des images de la face tout au long de l'essai de traction. Les images sont ensuite traitées à l'aide du logiciel de corrélation d'images *7D* [2], afin de calculer les déformations dans les 3 directions du repère lié à l'éprouvette. La courbe force-déplacement enregistrée par la machine de traction associée à une technique de recalage temporel permet d'obtenir la courbe contrainte-déformation et par suite d'en déduire certaines propriétés mécaniques élastiques, comme le module d'Young et les coefficients de Poisson du matériau. La mesure de ces paramètres nécessite de mesurer avec précision des très petites déformations (<0,05 %). Le challenge ici est donc d'exploiter au mieux les méthodes de corrélation d'image qui ont déjà fait leurs preuves pour la mesure de grandes déformations [3].

## 2. Matériau et procédé d'élaboration

Les matériaux composites ont été fabriqués à partir d'un renfort à fibres de lin et d'une résine époxyde SR 8100 associée au durcisseur SD 8822 (proportion en masse : 100/31). Se présentant sous la forme de rouleaux de tissus secs (figure 1a), le renfort est un sergé 2/2 à base de fibres de lin, de grammage 330 g/m$^2$ et de masse volumique 1450 kg/m$^3$ (selon les données du fournisseur). Un stratifié, constitué de 8 couches, a été réalisé au LISM (à l'IUT de Troyes) par le procédé de moulage par injection sous vide. Ce procédé consiste à mettre sous vide à l'aide d'une bâche plastique plusieurs couches de renforts secs empilées qui sont par la suite imprégnées de résine, l'imprégnation se faisant par aspiration créée par une dépression. Cette technique nécessite un filet de drainage pour solutionner les problèmes de flux de résine, un film séparateur perforé et un tissu d'arrachage. Après infusion, le cycle de polymérisation préconisé par le fabricant a été respecté, à savoir une 1$^{ère}$ réticulation à température ambiante pendant 24 heures suivie d'une 2$^{nde}$ pendant 16 heures à 60°C. La fraction volumique des fibres de lin de la plaque réalisée est de 30% ; celle-ci est calculée à partir des masses des fibres et de la plaque, de la masse volumique de la fibre de lin et de celle de la résine.

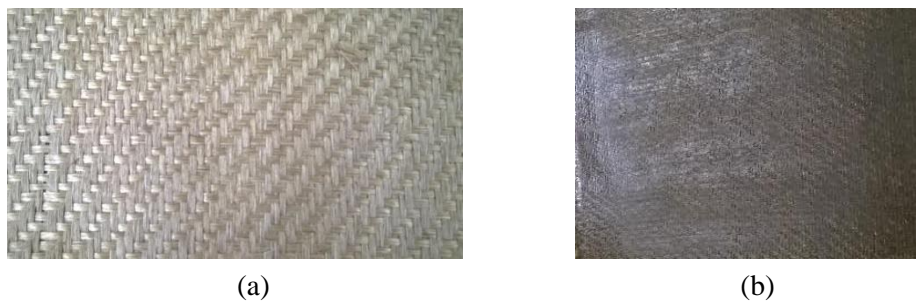

(a)                                            (b)

*Fig. 1. (a) Renfort sergé 2/2 à base de fibres de lin (b) Plaque de composite (8 plis)*

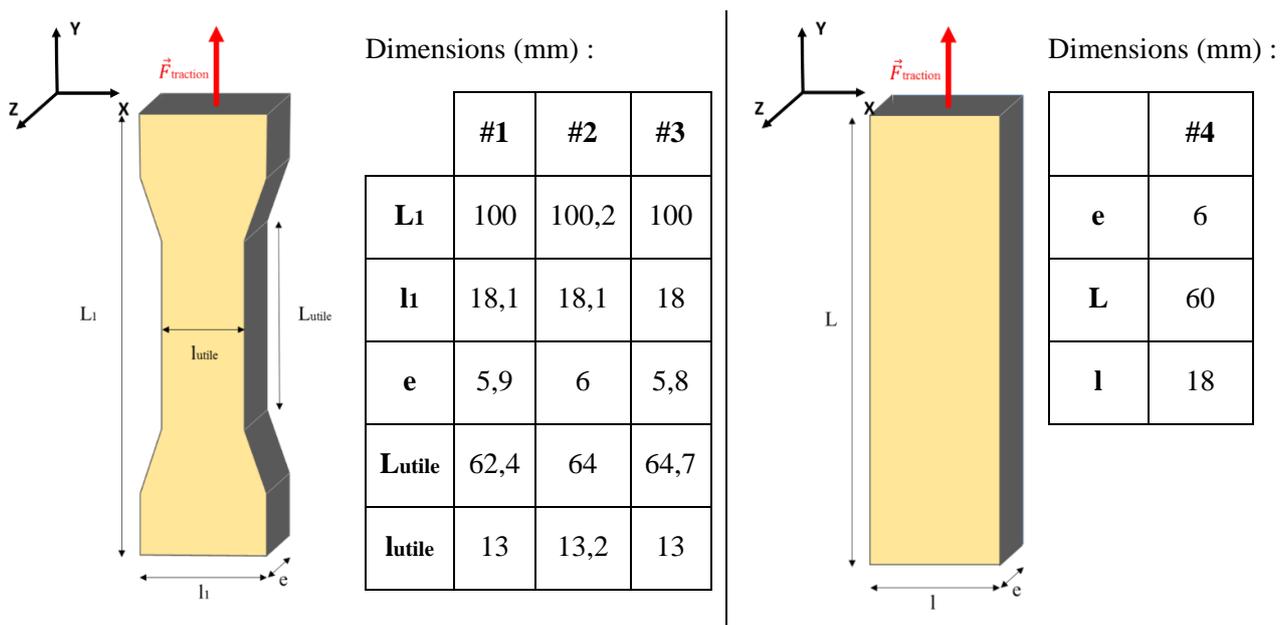

| Dimensions (mm) : | #1 | #2 | #3 |
|---|---|---|---|
| L$_1$ | 100 | 100,2 | 100 |
| l$_1$ | 18,1 | 18,1 | 18 |
| e | 5,9 | 6 | 5,8 |
| L$_{utile}$ | 62,4 | 64 | 64,7 |
| l$_{utile}$ | 13 | 13,2 | 13 |

| Dimensions (mm) : | #4 |
|---|---|
| e | 6 |
| L | 60 |
| l | 18 |

*Fig. 2. Formes et dimensions des éprouvettes*

Les éprouvettes sont ensuite découpées dans la plaque. Deux types d'éprouvettes ont été réalisés (figure 2) : des éprouvettes de forme haltère, conformément à la désignation D638M préconisée par la norme ASTM, et des éprouvettes rectangulaires. Les éprouvettes haltères sont usinées par contournage sur une fraiseuse tandis que les autres sont découpées avec une scie à fil.

## 3. Procédure expérimentale

### 3.1. Essais de traction

Les essais de traction sont réalisés sur une machine de traction INSTRON 5569 équipée d'une cellule de charge de 50 kN (précision : 0,5% de la charge) permettant d'enregistrer l'effort exercé sur l'éprouvette avec une fréquence d'acquisition de 5 Hz. Les trois grandeurs force, temps et déplacement de la traverse sont enregistrées par la machine au cours de l'essai de traction par le biais du logiciel Bluehill. Un extensomètre axial INSTRON 2620-604 a été utilisé pour mesurer les déformations longitudinales et confronter les mesures à celles obtenues par corrélation d'images. La figure 3 présente le dispositif.

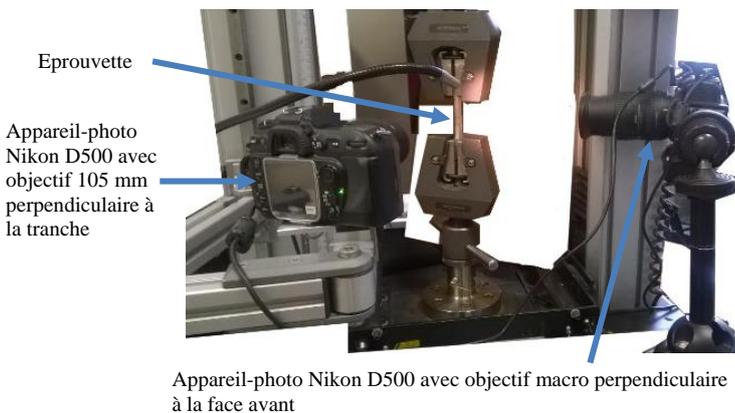

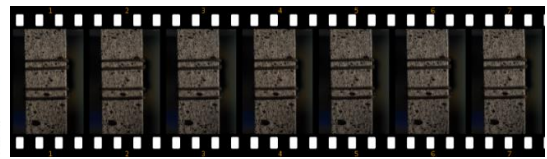

*Fig. 4. Séquence d'images au cours de la déformation de la face avant de l'échantillon (les bandes noires correspondent aux élastiques de maintien de l'extensomètre placé sur la face arrière de l'échantillon)*

*Fig. 3. Montage de l'essai de traction*

Pour chaque essai, une vitesse de traction $v$ = 0,6 mm/min correspondant à une vitesse de déformation $\dot{\varepsilon}$ = $1,67 \times 10^{-4}$ $s^{-1}$ a été retenue.

### 3.2. Corrélation d'images numériques

Deux appareil-photos Nikon D500, équipés respectivement d'un objectif macro et d'un objectif 105 mm, sont utilisés et enregistrent une séquence d'images à une fréquence d'acquisition définie au préalable (avec une limite de 0,5 Hz), les appareils étant synchronisés entre eux. Comme le montre la figure 3, l'un est placé de manière à avoir une vue plane de la face de l'éprouvette (objectif 105 mm) et l'autre permet d'avoir une vue plane de la tranche (objectif macro). Ils sont déclenchés en même temps que le départ de la machine de traction. Bien que la fréquence d'acquisition soit différente (le logiciel BlueHill enregistre les données toutes les 0,2s), une technique d'interpolation est mise en œuvre afin de faire correspondre les images avec les efforts mesurés lors de l'essai de traction. Par ailleurs, un éclairage composé d'une barre de lampes LED fait partie du montage et illumine l'échantillon, la qualité de l'éclairage étant très importante pour le post-traitement. En effet, l'image est plus contrastée, ce qui permet d'obtenir un spectre de couleurs plus vaste, ce qui n'est que plus avantageux avec *7D* puisque le logiciel travaille avec les niveaux de gris. La corrélation entre les images est ainsi plus facile puisque les motifs sont mieux reconnaissables.

Les séquences d'images (figure 4) sont analysées à l'aide du logiciel *7D*. La méthode d'analyse repose sur une technique de corrélation d'images numériques. Il s'agit d'une méthode permettant les mesures sans contact de la déformation des matériaux [4]. Sa mise en œuvre nécessite de recouvrir l'échantillon d'un mouchetis de peinture aléatoire au moyen de deux bombes de peinture. Dans le cas de cette étude, le mouchetis est noir et blanc. Le logiciel calcule les champs de déplacement entre deux images en utilisant la distribution de niveaux de gris au niveau des points concernés. Cette phase s'appelle l'appariement. Le logiciel divise la première image en une grille constituée de carrés dont la taille est à paramétrer au sein du logiciel. La taille optimale dépend de la dimension des taches sur le mouchetis. En effet, il ne faut pas qu'un carré se retrouve intégralement dans une tache car la zone où il est situé sera plus difficilement identifiée sur les autres images. Sur l'image suivante, le logiciel recherche le motif le plus semblable au motif initial pour chaque nœud (figure 5). Il est également possible de paramétrer la taille de la zone dans laquelle les motifs correspondants sont recherchés. Le calcul du champ de déplacements est effectué pour chaque image sélectionnée en utilisant une approche lagrangienne, c'est-à-dire basée sur l'image initiale non déformée. A l'issue des calculs, les champs de déplacements dans la zone analysée pour chacune des images et par suite les champs de déformations (logarithmiques ou Green-Lagrange) sont connus (figure 6).

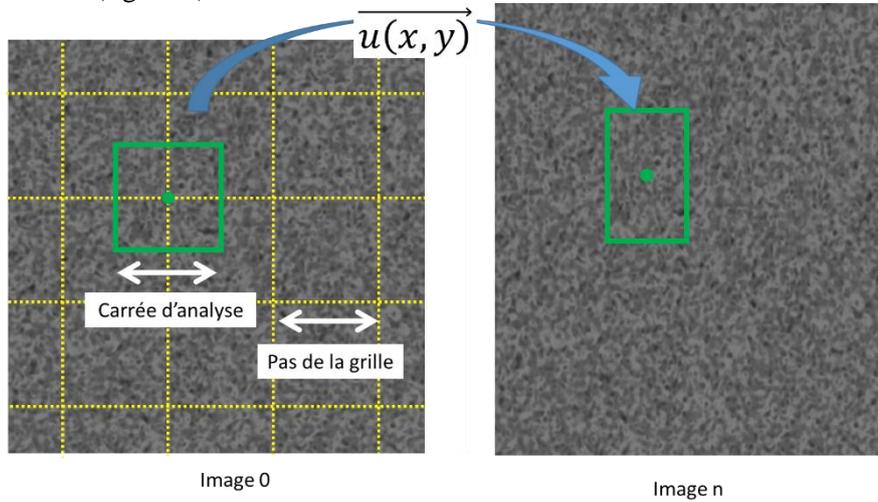

*Fig. 5. Principe de la corrélation d'images – définition du pas de la grille et de la taille du carré d'analyse*

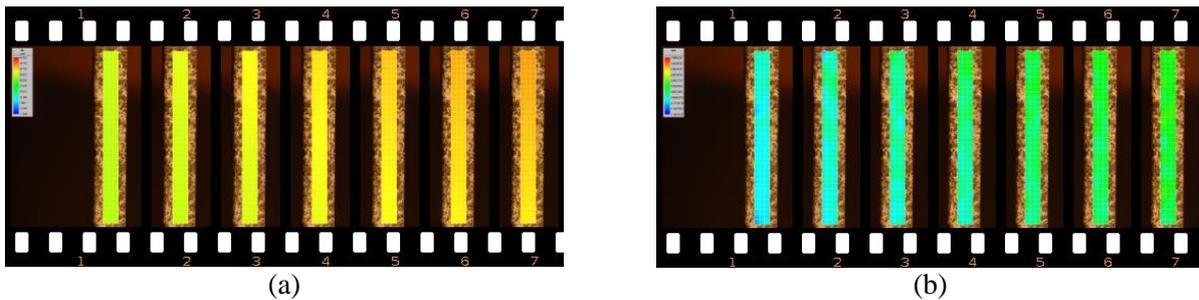

(a)  (b)

*Fig. 6. Séquence d'images de la tranche de l'éprouvette #3 : carte des champs de déplacement selon y (a) et des champs de déformation selon y (b)*

## 4. Résultats et discussion

### 4.1. Validation de la technique de corrélation d'images à l'aide d'un extensomètre

De par les très faibles niveaux de déformation dans le domaine élastique pour ce type de matériau (moins de 0,5%), la technique de corrélation d'images n'est pas à priori la technique la plus adaptée. Il s'agit tout de même de vérifier la validité des résultats obtenus par cette méthode en la confrontant à une seconde technique, plus courante : l'utilisation d'un extensomètre.

L'essai réalisé ici utilise ces deux techniques, ce qui permettra par la suite d'avoir la même référence pour les comparer. Un échantillon de forme rectangulaire est utilisé pour cet essai. L'extensomètre est fixé au moyen de deux élastiques sur la face arrière de l'échantillon et l'axe de l'objectif d'un appareil-photo est placé perpendiculairement à la face avant de l'échantillon. L'essai se déroule à une vitesse de 0,6 mm/min et les images sont prises à un intervalle de temps de 2s. Les déformations auxquelles l'extensomètre est soumis et la force de traction sont enregistrées à une fréquence d'acquisition de 5 Hz. L'essai est reproduit dans les mêmes conditions avec la même éprouvette afin de tester la robustesse de l'approche.

Pour déterminer les conditions optimales d'analyse par corrélation d'images, plusieurs analyses sont réalisées sur le même jeu d'images en modifiant les dimensions de la grille d'analyse et des carrés (figure 7 (a), (b) et (c)) et les zones d'analyse (figure 7 (d) et (e)).

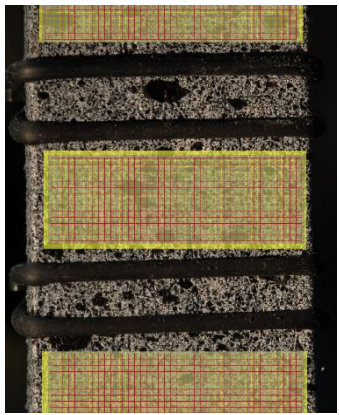
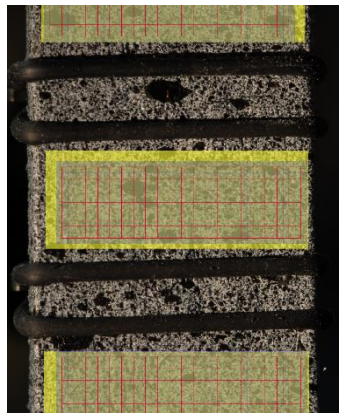
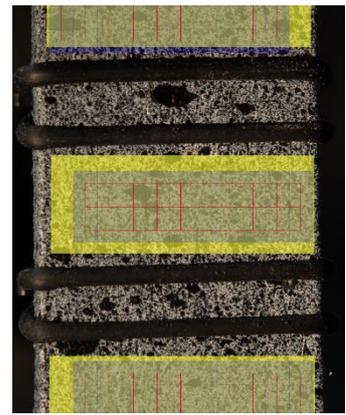

(a)  Pas de grille : 36x36 pixels
     Taille du motif : 36x36 pixels

(b)  Pas de grille : 64x64 pixels
     Taille du motif : 64x64 pixels

(c)  Pas de grille : 128x128 pixels
     Taille du motif : 128x128 pixels

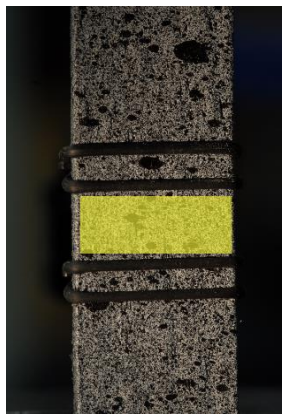
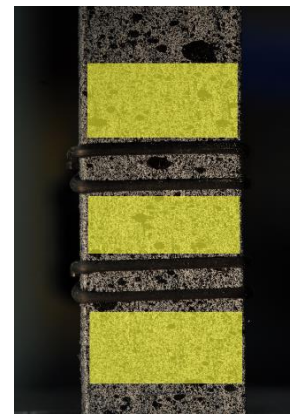

(d)  Zone d'analyse 1
(e)  Zone d'analyse 2

*Fig. 7. Dimensions des pas de grilles et des motifs et zones d'analyse*

La figure 8 présente la courbe contrainte nominale en fonction de la déformation longitudinale issue des analyses par corrélation d'images et de l'extensométrie classique. Les déformations inférieures à 0,001% issues de la corrélation d'images n'apparaissent pas sur cette figure car elles sont trop bruitées et donc peu exploitables. Les résultats obtenus par corrélation d'images sont comparables à ceux obtenus par extensométrie pour des déformations supérieures à 0,004%. Il est mis en évidence que la zone 2 fournit des résultats plus linéaires que ceux de la zone 1. On explique ces écarts entre l'extensomètre et les analyses d'images par le bruit induit par la présence des élastiques de l'extensomètre sur la surface d'analyse. Ce phénomène n'a pas lieu d'être en l'absence de l'extensomètre et on considèrera par la suite les résultats *7D* comme valides lorsque l'on choisit une taille de grille suffisamment grande ainsi qu'une zone d'analyse suffisamment large.

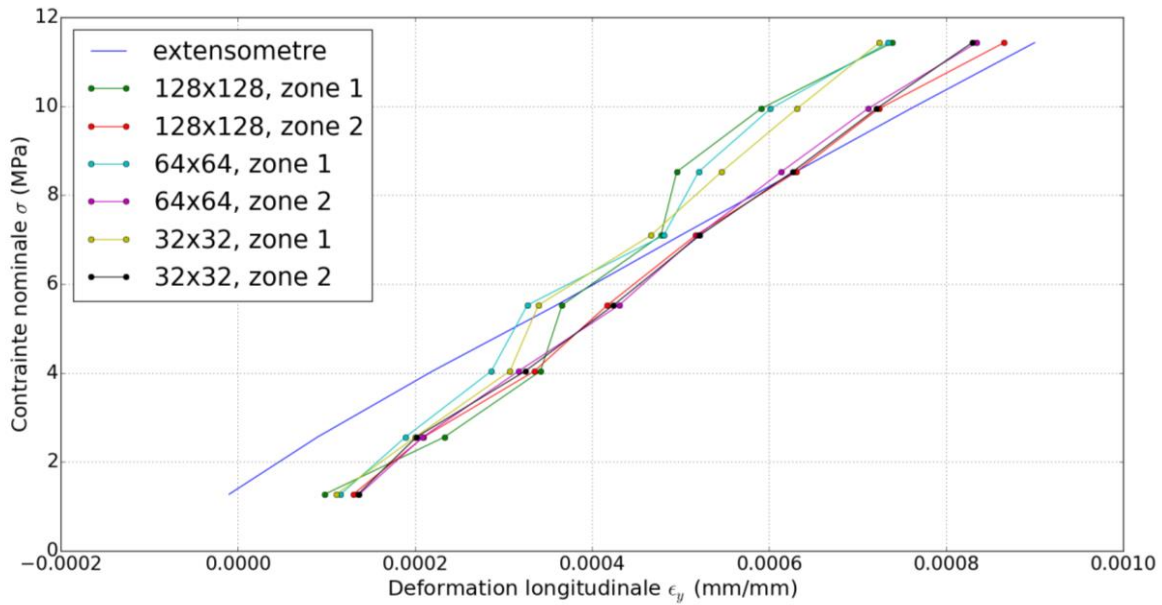

*Fig. 8. Tracé de la courbe contrainte nominale / déformation longitudinale avec les différents paramètres d'analyse*

### 4.2. Résultats des essais de traction

4.2.1. Contrainte à la rupture

La contrainte à la rupture est calculée à partir du rapport de la force à la rupture sur la section initiale de l'éprouvette (la section variant peu car les déformations sont très petites). Le tableau 1 donne les valeurs calculées pour les éprouvettes #1, #2 et #3 :

| **Eprouvette** | **#1** | **#2** | **#3** | **Moyenne** |
|---|---|---|---|---|
| **Contrainte à la rupture (MPa)** | 101,7 | 102,7 | 100,1 | 101,5 +/- 3 |

*Tab. 1. Contrainte à la rupture des trois éprouvettes*

En prenant en compte les incertitudes causées par les erreurs sur les mesures des sections des éprouvettes et de la charge, les résultats sont considérés comme précis à 3 MPa près. Les valeurs obtenues pour les 3 éprouvettes sont très proches puisque l'écart maximal entre contraintes à la rupture calculées est de 2,6 MPa. On peut donc considérer les résultats comme répétables. Finalement, la contrainte à la rupture moyenne est de 101,5 MPa, ce qui est en bon accord avec les valeurs rencontrées dans une précédente étude [1].

4.2.2. Module d'Young

Sous l'hypothèse d'un état de contrainte uni-axial, le module d'Young peut être calculé à partir de la loi de Hooke :

$$E = \frac{\sigma}{\varepsilon}$$

En pratique, ce dernier est calculé en traçant la contrainte nominale en fonction de la déformation longitudinale, la différence entre la contrainte nominale et la contrainte vraie étant négligeable au vu des résultats. Pour connaître la valeur de *E*, il suffit ensuite de déterminer la valeur du coefficient directeur de la courbe tracée à l'aide d'une régression linéaire.

La figure 9 trace la contrainte nominale en fonction des déformations longitudinales calculées par corrélation d'images avec les photos de la face de l'échantillon ainsi que celles de la tranche. Les résultats concordent bien même si les sources d'erreurs représentées par les carrés en pointillés sont nombreuses. Pour déterminer l'erreur sur les déformations, les écarts-types sont calculés. En ce qui concerne la contrainte, il faut prendre en compte les erreurs sur la valeur de la charge (0,5%), les erreurs de mesure de la largeur et l'épaisseur de l'échantillon et un éventuel décalage temporel entre les images et les résultats enregistrés par la machine de traction (au maximum 3s).

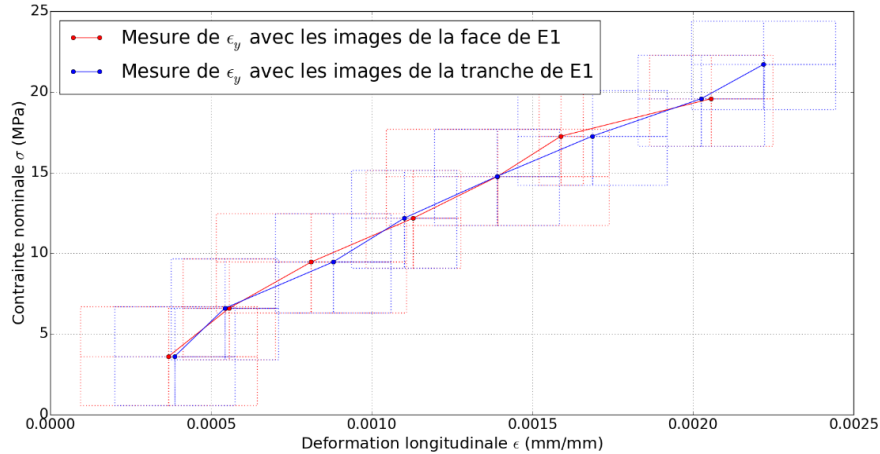

*Fig. 9. Contrainte nominale en fonction des déformations longitudinales calculées avec les images de la tranche et celles de la face pour l'échantillon 1*

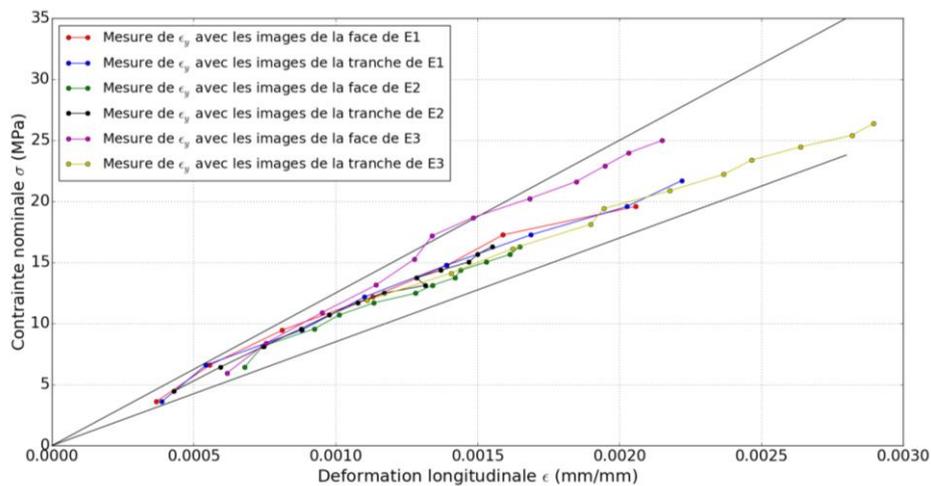

*Fig. 10. Module d'Young des éprouvettes 1 à 3 compris entre 8,5 et 12,5 GPa*

La figure 10 représente tous les résultats obtenus sur les éprouvettes 1, 2 et 3 et sont encadrés par deux droites de pentes respectives 8500 et 12500, ce qui donne un ordre d'idée de la valeur du module d'Young. Les résultats sont assez proches et hormis la mesure avec les images de la face de l'éprouvette #3, les résultats peuvent être considérés comme répétables.

D'autre part, à l'aide d'une régression linéaire, on obtient les valeurs de module d'Young suivantes avec les résultats de la corrélation d'images pour les éprouvettes #1 à #3 :

| Eprouvette | #1 | | #2 | | #3 | | Moyenne |
|---|---|---|---|---|---|---|---|
| Vue | Face | Tranche | Face | Tranche | Face | Tranche | |
| Module d'Young (MPa) | 8881 | 9425 | 9258 | 10127 | 12250 | 8580 | 9747 |

*Tab. 2. Module d'Young déterminé à partir de la corrélation d'images*

En comparaison, les résultats obtenus dans la partie 4.1 donnés par l'extensomètre avec l'éprouvette #4 sont les suivants : 11191 MPa pour le premier essai et 10212 MPa avec le second essai et corroborent bien les résultats obtenus par corrélation d'images.

4.2.3. Coefficients de Poisson

La corrélation d'images numériques présente des avantages par rapport à l'extensométrie classique puisqu'elle permet de déterminer l'ensemble des champs de déplacements et déformations dans les directions longitudinales et transverses tandis que l'usage d'un extensomètre seul n'apporte d'informations que pour une direction. De plus, l'utilisation de deux appareil-photos permet d'avoir ces données à la fois pour la tranche et pour la face de l'éprouvette. On va ainsi pouvoir obtenir les valeurs des coefficients de Poisson dans le plan de la face de tissage de l'échantillon ($\nu_{yx}$) et également dans le plan de l'épaisseur du tissage ($\nu_{yz}$). Les coefficients de Poisson sont calculés à l'aide des formules suivantes, en considérant la direction de traction suivant l'axe *y* :

$$\nu_{yx} = -\frac{\varepsilon_x}{\varepsilon_y} \; et \; \nu_{yz} = -\frac{\varepsilon_z}{\varepsilon_y}$$

En pratique, pour chaque image, une carte de déformations est obtenue (la déformation est mesurée en chaque nœud de la grille d'analyse). Cette information locale est ensuite moyennée sur toute la zone d'analyse de l'image, ce qui permet de connaître la déformation moyenne dans la direction de traction et dans la direction transverse pour chaque image. Il est à noter que cette méthode, par la moyenne, est en accord avec l'hypothèse de base : la déformation est constante dans la zone utile de l'éprouvette. Les deux déformations moyennes pour chaque image sont ensuite reportées sur un graphe, puis par régression linéaire, le coefficient de Poisson est obtenu (au signe près).

| Eprouvette | #1 | | #2 | | #3 | |
|---|---|---|---|---|---|---|
| Coefficient de Poisson | $\nu_{yx}$ | $\nu_{yz}$ | $\nu_{yx}$ | $\nu_{yz}$ | $\nu_{yx}$ | $\nu_{yz}$ |
| | 0,10 | 0,33 | - | 0,75 | 0,42 | 0,33 |

*Tab. 3. Coefficients de Poisson déterminés à partir de la corrélation d'images*

Comme le montre le tableau 3, les premiers résultats obtenus pour le calcul des coefficients de Poisson par corrélation d'images ne sont pas probants et la technique employée nécessite encore d'être affinée. En effet, les valeurs obtenues pour les coefficients de Poisson divergent beaucoup, hormis pour ceux des tranches des éprouvettes 1 et 3, et certaines sont incohérentes. Ces divergences et incohérences sont dues au fait que les déformations dans les directions transverses sont très petites et sont donc difficiles à calculer avec la méthode de corrélation d'images utilisée dans cette étude. Il est indispensable d'approfondir l'approche en utilisant d'autres méthodes, notamment la méthode intégrée [5], qui peut résoudre ce problème.

## 5. Conclusion

La méthode, présentée ici, pour mesurer les paramètres anisotropes de la loi élastique pour un matériau composite à fibres de lin et d'une résine époxyde est très prometteuse. La mesure notamment des propriétés élastiques dans l'épaisseur du matériau est rendue possible grâce à la corrélation d'images numériques.

Pour l'estimation du module d'Young longitudinal et de la contrainte à la rupture, les valeurs moyennes obtenues par cette technique sont en accord avec celles de l'extensométrie classique.

Malheureusement, la même conclusion ne peut pas être émise pour l'instant quand il s'agit de déterminer les coefficients de Poisson. Il faut en effet noter que les déformations à mesurer sont très petites et sont à la limite des capacités des méthodes conventionnelles de corrélation d'images. L'utilisation de la méthode intégrée, où le champ de déplacement (linéaire dans ce cas) est ajouté comme contrainte dans l'algorithme de corrélation d'images pourra permettre d'aller plus loin dans la démarche et de repousser la limite basse des déformations mesurées.